\documentstyle[aps,prl,preprint]{revtex}
\begin{document}
\draft
\tighten
\title{Two-pion exchange potential and the $\pi N$ amplitude}

\author{M. T. Pe\~na}
\address{Centro de F\'{\i}sica Nuclear da
Universidade de Lisboa, 1699 Lisboa Codex\\
and
Instituto Superior T\'ecnico, 1096 Lisboa Codex}

\author{Franz Gross and Yohanes Surya}
\address{College of William and Mary, Williamsburg, VA 23185\\
and
CEBAF 12000 Jefferson Avenue, Newport News, VA 23606}

\maketitle

\begin{abstract}

We discuss the two-pion exchange potential which emerges from a
box diagram with one nucleon (the spectator) restricted to its mass shell,
and the other nucleon line replaced by a subtracted, covariant $\pi N$
scattering amplitude which includes
$\Delta$, Roper, and $D_{13}$ isobars, as well as contact terms
and off-shell (non-pole) dressed nucleon terms. The $\pi N$ amplitude
satisfies chiral symmetry constraints and fits $\pi N$  data below $\sim$
700 MeV pion energy. We find that this TPE potential
can be well approximated by the exchange of an effective sigma and delta
meson, with parameters close to the ones used in one-boson-exchange
models that fit $NN$ data below the pion production threshold.

\end{abstract}
\narrowtext

\section*{Overview and Framework}

At low energies it seems to be possible to describe hadronic processes
in terms of nucleons
and mesons alone. Effective dynamical models for the $NN$ interaction can
be constructed with these underlying degrees of freedom.
Meson exchange models are then favored since
they allow a straightforward consistent connection
to the electromagnetic and electroweak reactions.
However, to extend these models up to a few GeV lab energy, as needed for 
the study of many reactions produced at
CEBAF and other modern accelerators, one must include many inelastic
channels associated with the explicit production of pions (and other mesons).
These channels are often described by baryon isobars, including their 
decays and propagation.  

At these higher energies, one must also include
relativistic effects which arise not only from the kinematics and the need
to correctly transform amplitudes from one frame to another, but also from
the nonlocallity and energy dependence which characterize relativistic 
interactions obtained from an effective field theory.  The correct treatment
of unitarity also requires that the the renormalization
of the meson-nucleon vertices and the simultaneous dressing of the ``bare'' 
nucleons by meson rescattering be consistent with the meson exchange
picture of the nuclear force\cite{Sa}.  It is difficult to deal
consistently with this complex problem  within strictly
non-relativistic formalisms\cite{Bl}, and we will show in this paper that  
a relativistic effective field theory provides a natural way to attack
this problem.

Using the covariant spectator (or Gross) equations\cite{Gross}, 
relativistic one boson exchange (OBE) models 
have been found which give an excellent description of elastic $NN$ 
scattering (ie. below the pion production threshold)\cite{GrOr}.  It is the
purpose of this paper to outline how this program should be extended
to include the inelastic effects arising from the explicit production
of pions in intermediate states.  We will limit ourselves to consideration
of the lowest lying inelastic cut arising from the $\pi NN$ channel,
and postpone any discussion of how to treat the $n\pi NN$ channels 
(with $n>1$).

Consider a simple model with a nucleon ($N$) and its first excited 
state ($N^*$) coupled to a pion.  Assume that 
the two baryons $N$ and $N^*$ are both bound states of some other fundamental
constituents confined inside the baryon volume, so that in
the absence of the interaction with the pion they are both stable states.  
The coupling to the pion then
dresses the $N$ and, if we assume $m_{_{N^*}}>m_{_N}+\pi$, allows the
$N^*$ to decay to $N+\pi$.  To treat the
$\pi NN$ cuts consistently in a {\it time-ordered 
formalism\/}, it is then sufficient to ignore diagrams which contribute 
to $2\pi NN$ cuts, such as those with
$\pi NN^*$ and $N^*N^*$ channels.  Examples of classes of diagrams 
which must be 
included and classes which can be ignored are shown in Fig.~1.  The
diagrams in Figs.~1a are iterations of bare $N$ and $N^*$ states, 
the $N^*N^*$ channel being excluded because its cuts are too high.
Diagrams in Figs.~1b are generated from those in 1a through the consistent
application of unitarity.  They include nucleon dressing, the
virtual transition of $N^*\to N$ (when not forbidden by selection
rules), and the transitions $N^*\to \pi N\to N^*$, which give the
$N^*$ its width in this model.  Diagrams in Figs.~1c all have 
higher cuts and are ignored in this approximation.  Note that this
eliminates the need to consider the $\pi N^*N^*$ vertex and the 
$\pi N^*$ dressing of the baryons.

The time ordered formalism is not covariant, and it is
desirable to find diagrams from covariant perturbation theory which
include the physics contained in Figs.~1a and b.  
Motivated by the advantages of the Gross or "spectator" prescription
for dealing with the relative energy variable in four-dimensional
integral equations \cite{Gross}, we chose the class of diagrams 
shown in Fig.~2.  In these figures the ``X'' means that the nucleon is
on-shell; which means that  its propagator is replaced by
\begin{equation}
{\not p + m_{_N}\over m_{_N}^2-p^2} \to \delta_+(m_{_N}^2-p^2)\;
\sum_s u(p,s)\overline{u}(p,s) \, . \label{eq1}
\end{equation}
Restricting the spectator nucleons in $NN^*$ and $N\{\pi N\}$ loops
to their mass-shell does not alter the $\pi NN$ 
unitarity cuts contained in these diagrams.
This choice is also consistent with our aim
to relate the present work to the one of Ref.~\cite{GrOr}, and differs from
the choice we would have made if we wished to extend
Refs.~\cite{Pan,Tjon,Elster,Riska}.   In Ref.~\cite{Pan} a non-relativistic
equation is used, in Ref.~\cite{Tjon} the Bethe-Salpeter equation is solved
in ladder approximation, and in  Refs.~\cite{Elster,Riska} the Blankbecler
and Sugar reduction is used.  Each of these approaches would suggest a
particular way of handling the diagrams in Fig.~1.  Since the Gross equation
specifies that one particle is on the mass-shell in any intermediate state,
the covariant class of diagrams  shown in Fig.~2, with the substitution of
Eq.~(\ref{eq1}), is the correct way to incorporate the physics of Fig.~1 into
that formalism.  

Our discussion implies that the full treatment of the coupled $NN - 
\pi NN$ system requires the summation of the class of diagrams shown,
up to 6th order, in Fig.~3.  A general discussion 
of the coupled equations which sum this infinite class of diagrams 
will be presented elsewhere.  Here we will limit 
ourselves to an evaluation of the kernel shown in Fig.~4.  When this
kernel is added to the OBE kernel used in the relativistic treatment
of the $NN$ channel below the
pion production threshold \cite{GrOr}, we have an equation which sums the 
subclass of
diagrams shown in Fig.~3a, but omits contributions of the type illustrated
in Fig.~3b.  Study of the kernel shown in in Fig.~4 will be a first 
estimate of the effect of $\pi N$ rescattering in the spectator
formalism.

This diagram is evaluated by carrying out the three-dimensional numerical
integration over the internal three-momentum because the spectator
prescription of Eq.~(\ref{eq1}) specifies the internal energy. Furthermore,
since the iteration of the one pion exchange (OPE) graphs with bare $NN$
intermediate states are already included when the $NN$ scattering is
described using the OBE kernel, the contribution from the undressed direct
nucleon pole (with bare Dirac nucleon propagator describing the intermediate
nucleon) must be subtracted from the
$\pi N$ amplitude which is used in Fig.~4 in order to avoid double counting.
The dressing of the nucleon pole term which arises from $\pi N$ rescattering
must be retained. In a short-hand
symbolic notation, the kernel obtained from Fig.~4 is then
\begin{equation}
V_{\rm{TPE}}=M^{4}-\int d^3k\; V_{\pi} \;{\cal G}\; V_{\pi}\, , \label{eq2}
\end{equation}
where $M^{4}$ is a diagram similar to Fig.~4 but with the ``blob'' replaced
by the {\it full\/} $\pi N$ scattering amplitude, $V_{\pi}$ is the OPE kernel,
and $\cal{G}$ the symmetrized sum of one undressed Dirac propagator and one
mass-shell projection operator as given in Eq.~(\ref{eq1}).  Subtracting the
iterated OBE means that (\ref{eq2}) can be added to the OBE kernel of 
Ref.~\cite{GrOr} without double counting. 

Before turning to the detailed study of the two pion exchange (TPE)
kernel, we briefly address a question of consistency which has posed
a serious problem in previous studies of pion production.  Consider
the diagram shown in Fig.~2b1.  In this diagram, one of the nucleons
is off-shell and is being dressed by the $\pi N$ interaction while the 
other is a physical on-shell nucleon.  How can these two nucleons be
considered identical when they are treated in such a different way?
This question has a very simple answer in the spectator formalism.
First, we must symmetrize the interaction, and this is the reason why
both diagrams shown in Fig.~4 must be included in the kernel.  If, in
addition,  
we require that the {\it properties of a dressed nucleon be identical
to a free nucleon when it is on mass-shell\/}, then the two nucleons will
be identical in every respect.  The model we will
use to describe the $\pi N$ scattering part\cite{GrSu} of the TPE kernel 
has precisely this feature.
  
\section*{The TPE Kernel}       

As discussed above, we limit ourselves in this work to a calculation of the
TPE potential given in Eq.~(\ref{eq2}) and shown
diagramatically in Fig.~4, in which the full $\pi N$ amplitude (with the {\it
direct nucleon pole term removed\/}) is embedded in the two-nucleon system. 
As we have emphasized above, it is essential to remove the direct nucleon pole
term from the $\pi N$ scattering amplitude (so the ``blob'' in
Fig.~4 represents this difference) in order to avoid counting the nucleon box
contribution more than once. 

According to the conventional wisdom underlying the one boson exchange model,
the process of Fig.~4 contributes to the intermediate and short range
behavior of the $NN$ interaction, and is represented by the exchange of
effective bosons with masses heavier than two pion masses (ie. greater than
$\simeq 280$ MeV).  Such a representation cannot be valid above the
$\pi NN$ threshold, however, where production of a physical $\pi NN$
intermediate state ensures that the $NN$ scattering will be
inelastic\cite{Elster}.  Our calculation will be applied below the
pion production threshold, and is only meant to be a first step to
estimating the contributions from nucleon resonances and to preparing the way
for the extension of the models of Ref.~\cite{GrOr} to higher
energies.  We will conclude from the results that the diagram of Fig.~4 
gives a  good
explanation of the appearance of effective $\sigma$ and $\delta$ mesons, 
which describe
the TPE exchange but do not correspond to real mesons. 

The calculation of the diagram shown in Fig.~4 is greatly simplified if we
use a $\pi N$ amplitude with a simple, manifestly covariant sructure. We take
here the covariant $\pi N$ amplitude already constructed in 
Ref.~\cite{GrSu} and  applied
successfully (with minor changes) to the description of pion 
photoproduction \cite{SuGr} and
to the calculation of the $\Delta$ isobar $E_2 / M_1$ ratio.  The covariant 
$\pi N$
amplitude is decomposed into contributions even and odd under the exchange 
of isospin
\begin{equation}
T=T^+ \delta_{ij} + T^- [{\tau}_j \, , \,{\tau}_i] 
\end{equation}
where each component depends only on the total four-momentum of the $\pi
N$ system, which we denote by $P$,   
\begin{equation}
 T^{\pm}(P^2) = A^{\pm}(P^2) + B^{\pm}(P^2) \rlap/ P \, .  
\end{equation}
We use the latest version of the model given in Ref.~\cite{SuGr}.
Details can be found both in both \cite{GrSu} and \cite{SuGr},
where a description of the phase shifts and inelasticities 
is also given.

This calculation of the effective TPE potential differs substantially from
previous work.  Most other calculations have limited contributions to
the $\pi N$ amplitude which appears in Fig.~4 to the $\Delta$ isobar
\cite{Pan,Tjon,Elster}.  Some have used dispersion relations to include other
resonances \cite{Riska}.  Our calculation includes the following features
\begin{itemize}
\item the $\pi N$ amplitude is obtained beyond
the tree-level, by
solving a relativistic two-body scattering integral equation where
the pion is considered on its mass-shell in all the intermediate
states;
\item  the kernel of the $\pi N$ equation includes the nucleon, $\Delta$,
Roper
and $D_{13}$ poles together with a contact term which includes effects from
the crossed nucleon 
pole and $\sigma$ and $\rho$
exchange terms, essential to satisfy chiral symmetry constraints and include
the physics of the inelastic $\pi NN$ channel approximately;
\item the spin $3/2$ particles ($\Delta$ and $D_{13}$) are described
by a propagator that includes a covariant spin $3/2$ projection
operator with the pole at $P^2=0$ cancelled by a zero in the baryon form
factor;
\item the baryon form factors vanish not only at at $P^2=0$, but also 
in the space-like region $ P^2 < 0$ (which lies far away from the
physical scattering region).
\end{itemize}

\noindent Since the amplitude contains a description of the crossed
baryon poles, the crossed-box diagram shown in Fig.~5 is automatically
included, although in an approximate fashion.

The overall isospin factors for the isospin $1/2$ and $3/2$ contributions to
the $\pi N$ amplitudes are ($P^{ij}_{1/2}$ and $P^{ij}_{3/2}$ being the
isospin projection operators for the $1/2$ and $3/2$ cases)
\begin{eqnarray}
{(\tau_j \tau_i)}_1 {(P^{ij}_{1/2})}_2 &&= {(P^{ij}_{1/2})}_1
{(\tau_j \tau_i)}_2 = {1\over3}{(\tau_j
\tau_i)}_1 {(\tau_j \tau_i)}_2= 1-{2\over3}{\bf \tau}_1  \cdot 
{\bf \tau}_2 \nonumber\\
{(\tau_j \tau_i)}_1 {(P^{ij}_{3/2})}_2 && = {(P^{ij}_{3/2})}_1{(\tau_j
\tau_i)}_2 = {(\tau_j \tau_i)}_1\left[\delta_{ij} -{(P^{ij}_{1/2})}_2\right] 
= 2+{2\over3}{\bf \tau}_1  \cdot {\bf \tau}_2
\end{eqnarray}
This very simple argument shows that the strength of the exchange
of an effective isoscalar meson is the sum of contributions from both $I=1/2$
and $I=3/2$ $\pi N$ channels (weighted 1 to 2), while the strength of the
exchange of an effective isovector meson is the {\it difference\/} of the
$I=1/2$ and $I=3/2$ contributions. In a model dominated by the $\Delta$
isobar, we expect the effective strengths of the isoscalar and
isovector exchanges to be in the ratio of three to one, and to have the
same sign.  However, when the $I=1/2$ contributions to the internal $\pi N$
scattering are important (which is true for the $D_{13}$ contribution),
the relative strengths of the different effective exchanges can be very
different.  Our calculation shows that the relative sizes of the $\Delta$
and $D_{13}$ amplitudes are such that, in the isovector exchange channel,
there is an almost perfect (and unexpected) cancellation between the two,
leaving a very small (but not zero) effective isovector exchange force. 
This is probably the most surprising result of this work. Signs
of this cancellation were present in Ref.~\cite{Riska}, although it was not
seen as spectacularly, since in that paper the
$D_{13}$ resonance was modeled with a width of roughly 1/4
of its present experimental value. Yet, in that calculation, the
$D_{13}$ resonance
was already found to be needed to oppose the contribution of the
$\Delta$ to the short range central part of the NN potential.

\section*{Results and Conclusions}

If all of the external nucleons are on-shell, the TPE exchange amplitude shown
in Fig.~4 consists of six independent helicity
amplitudes for each possible value of the isospin ($I=0$ or $I=1$) in the
$t$ (exchange) channel.  (For a full discussion of the definition of $NN$
helicity amplitudes see, for example, Ref.~\cite{GrOr}.)  All of these
amplitudes were evaluated and compared to those which are obtained from the
exchange of a scalar meson (either $\sigma$, if $I=0$, or $\delta$, if $I=1$)
with an arbitrary mass and coupling constant, and a meson$NN$ form factor
with a cutoff mass around $1$--$1.2$ GeV. The form factor mass was held
constant during each fit, and was chosen to most closely match the energy
dependence of the TPE amplitude.  Once the form factor mass was chosen, the
meson mass and coupling constant were adjusted to give the best fit.

Figure 6 shows the quality of the fit for several helicity amplitudes at two
different relative momenta $P$ of the $NN$ pair, both corresponding to an
energy below the one pion production threshold (which is at $P=370$ MeV/c). 
Note that the fits are quite good, showing that these TPE potentials are well
represented by a sum of $\sigma$ and $\delta$ exchanges (at least below the
pion production threshold).  For each of the isoscalar and isovector channels
we also tried fitting to the sum of scalar plus vector exchange terms, which
for the $I=0$ channel included $\sigma$ and $\omega$ exchanges, and for the
$I=1$ channel included $\delta$ and $\rho$ exchanges.  We found that the fits
did not significantly improve, showing that the TPE kernel provides little
justification for the addition of effective vector mesons to our
effective $NN$ force.  In fact, the deviation from zero of the differences
$M^4_{++-+} - M^4_{+-++}$ and $M^4_{++++} - M^4_{++--}$ cannot be explained
through a $\sigma$ ($\delta$) exchange alone, but these differences, which
increase slowly with energy, were found to be very small.  This result
confirms the observation that the full TPE amplitudes can be approximated by
the exchange of scalar mesons only (at least below the pion production
threshold).

The effective meson parameters resulting from the fits are shown in Table
I. In both the isoscalar and isovector cases the values obtained for the mass 
and coupling constant are somewhat smaller than the values obtained from OBE
models, but are still in reasonable agreement with those results.

Figure 7 shows how a typical $NN$ amplitude ($M^4_{++++}$) is built up from
the different parts of the $\pi N$ amplitude, and also displays the
cancellation of the $\Delta$ and the $D_{13}$ contributions in the isovector
channel.  It also shows that the $\Delta$ contribution dominates the
isoscalar exchange channel.

We conclude that the large effective $\sigma$ exchange, together
with the comparatively small effective $\delta$ exchange, both of
which are characteristic features of OBE models, can be largely explained by
the TPE contribution, {\it provided the contribution of the $D_{13}$
resonance is properly included\/}.

\acknowledgements

M.~T.~Pe\~na thanks the CEBAF Theory Group for its hospitality during her
extended visits to CEBAF. The work of M.~T.~Pe\~na is supported by JNICT, 
under Contract numbers No.\ PBIC/C/CEN/1094/92 and
No.\ CERN/P/FAE/1047/95.  The work of F.~Gross is supported by DOE Grant
No.~DE-FG05-88ER40435.

\nopagebreak[4]

\begin{table}
\begin{tabular}{ccccc}
  & mass (IIB) &mass (TPE) & $g^2/(4\pi)$ (IIB) & $g^2/(4\pi)$ (TPE)\\
\tableline
 $\sigma$ & 522 & 508  & 4.87 &  4.02  \\
 $\delta$ & 428 & 345  & 0.24 &  0.20
\label{tab1}
\end{tabular}
\caption{Comparison of the sigma and delta masses and couplings obtained for
Model IIB of Ref.~[4] with the same parameters obtained from a fit to
the TPE helicity amplitudes.  Masses are in MeV; the $\delta$ meson was
denoted by $\sigma_1$ in Ref.~[4].}
\end{table}

\vspace*{10mm}
\begin{figure}
\caption{Time ordered diagrams describing pion production and rescattering.
(a) Pion exchange, including resonance contributions; (b) self energy
diagrams generated by unitarity from the same cuts shown in (a); (c)
diagrams which can be ignored because they have cuts above the $2\pi NN$
production threshold.}
\vspace*{0.2in}
\caption{Covariant diagrams for use in the spectator formalism which
include the physics of the diagrams shown in Fig.~1a and 1b.}
\vspace*{0.2in}
\caption{All diagrams up to 6th order in the $\pi NN$ coupling which are
iterations of the basic interactions included in Fig.~2.}
\vspace*{0.2in}
\caption{The effective TPE amplitude studied in this paper.  The ``blob'' is
the full $\pi N$ amplitude with the direct nucleon pole term removed.}
\vspace*{0.2in}
\caption{ the crossed pion exchange diagram included in the TPE kernel.}
\vspace*{0.2in}
\caption{Isoscalar and isovector TPE helicity
amplitudes, at relative $NN$ momenta of 60 MeV/c and 240 MeV/c, compared to
the fitted sigma ($I=0$) and delta ($I=1$) one boson exchange amplitudes. }
\vspace*{0.2in}
\caption{Contributions of individual parts of the $\pi N$ amplitudes to the
$V_{{\rm TPE}}^{++++}$ amplitude.  The left two pannels are the isoscalar
exchange amplitude, the right two are isovector, the top two are for
$P=60$ MeV, and the bottom two for $P=240$ MeV.  Note that the isoscalar
amplitudes are well approximated by the $\Delta$ contribution alone, while
the $D_{13}$ and $\Delta$ contributions cancel for the isovector terms.}
\end{figure}

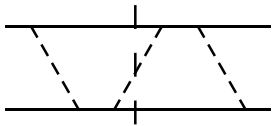
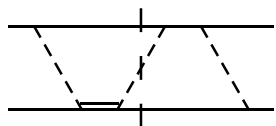
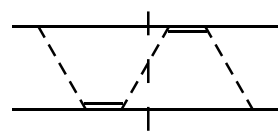
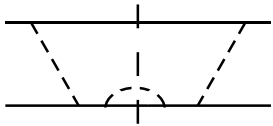
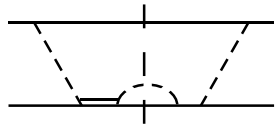
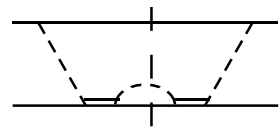
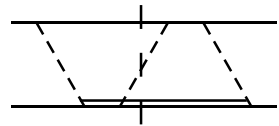
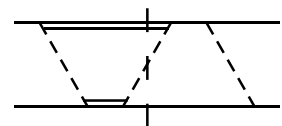
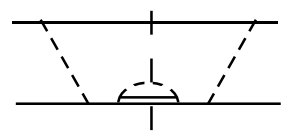
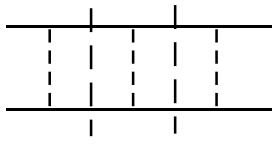
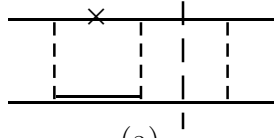
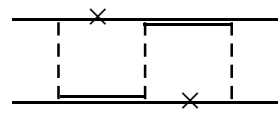
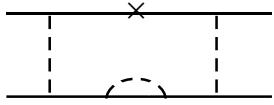
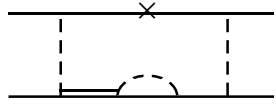
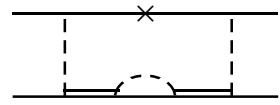
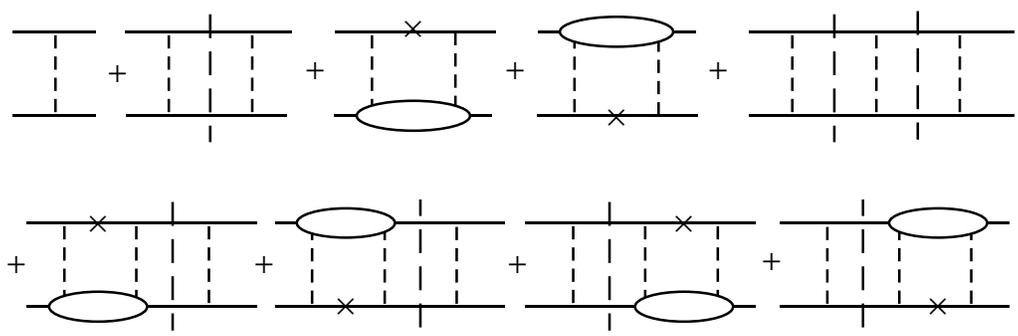
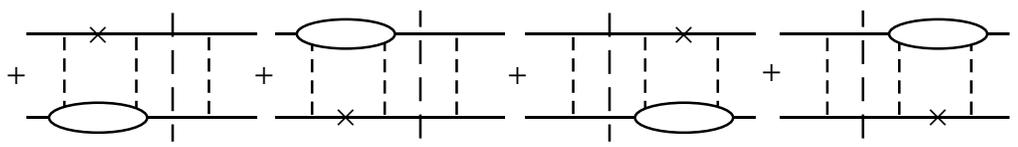
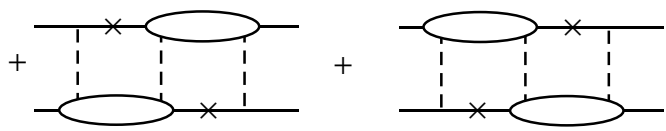
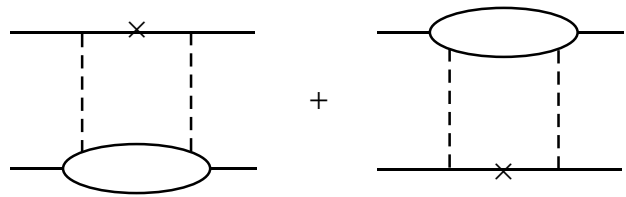
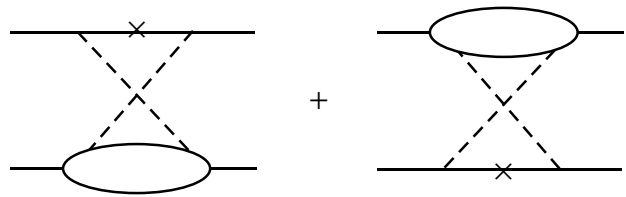
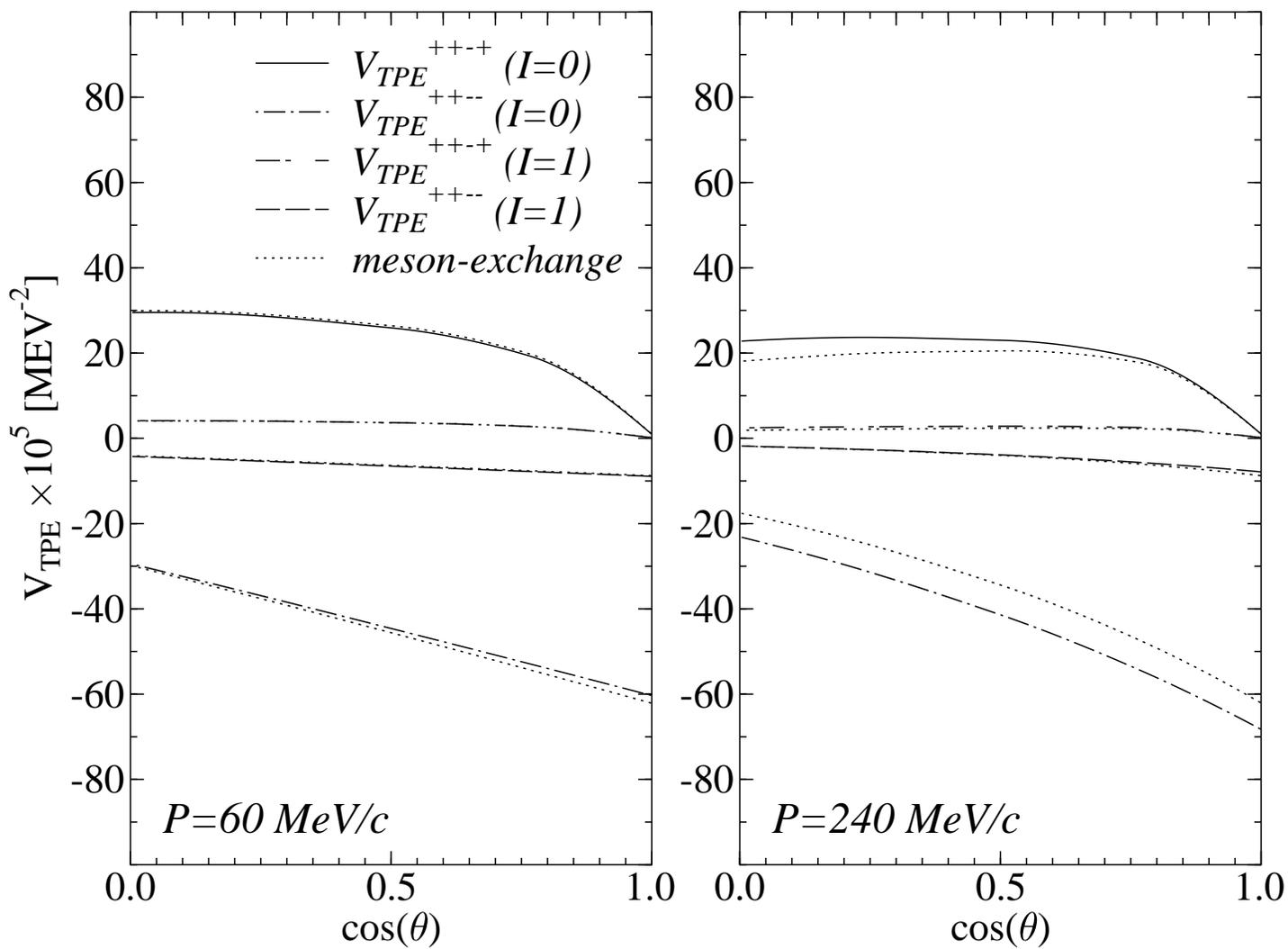
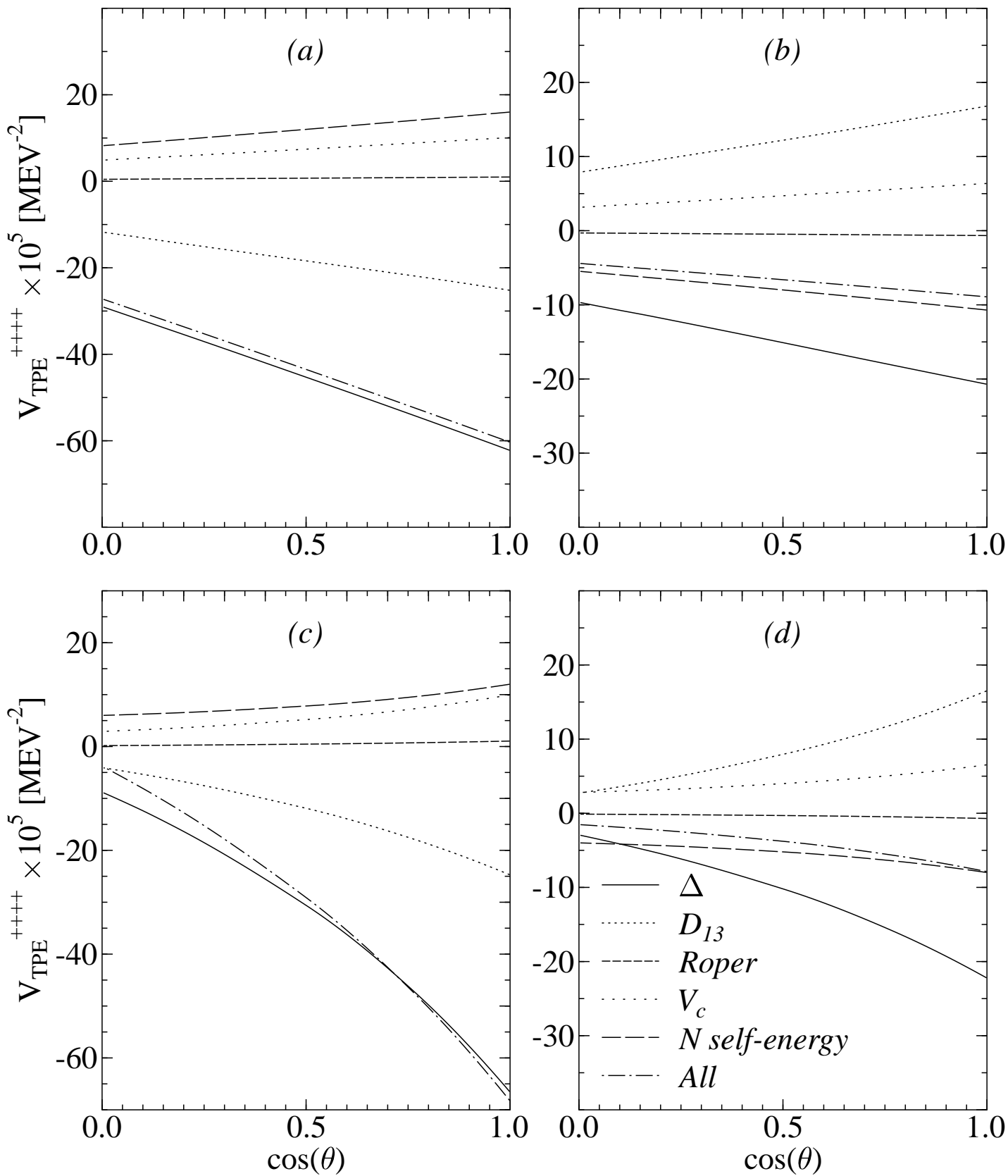
\end{document}